\documentstyle[preprint,aps]{revtex}
\tighten

\begin{document}
\draft

\newcommand{\be}{\begin{eqnarray}}
\newcommand{\ee}{\end{eqnarray}}
\newcommand{\beq}{\begin{equation}}
\newcommand{\eeq}{\end{equation}}
\def\ni{\noindent}
\def\ul{\underline}
\def\in{\indent}
\def\lin{\in\in\in}
\def\sss{\scriptscriptstyle}

\title{The Spectral Line Shape of Exotic Nuclei}

\author{
F. Ghielmetti$^{1}$, G. Col\`o$^{1,2}$, E. Vigezzi$^{1,2}$, 
P. F. Bortignon$^{1,2}$, R. A. Broglia$^{1,2,3}$
}

\address{$^1$ Dipartimento di Fisica, Universit\`a di Milano,\\
Via Celoria 16, 20133 Milano, Italy.\\
$^2$ INFN Sezione di Milano,\\
Via Celoria 16, 20133 Milano, Italy.\\
$^3$The Niels Bohr Institute, University of Copenhagen,\\
2100 Copenhagen, Denmark.
}

\vspace{1cm}

\date{\today}

\maketitle

\begin{abstract}
The quadrupole strength function of $^{28}O$ is calculated making use of the 
SIII interaction, within the framework of continuum-RPA 
and taking into account collisions among the nucleons (doorway
coupling). The centroid of the giant resonance is predicted 
at $\approx 14$ MeV, that is much below the energy expected for both 
isoscalar and isovector quadrupole resonances in  
nuclei along the stability valley. 
About half of this width arises from the coupling of the resonance to the
continuum and about half is due to doorway coupling. This result is similar
to that obtained in the study of giant resonances in light, $\beta$-stable
nuclei, and shows the lack of basis for the expectation, entertained until
now in the literature, that continuum decay was the main damping mechanism
of giant resonances in halo nuclei\ \cite{sug_ref}. 
\end{abstract}

\narrowtext
 
\newpage 

A central subject in the study of the nuclear structure is that of the 
damping of vibrational motion (cf., e..g.,\ \cite{Ber83,Ber94}). 
The variety of relaxation processes 
may be divided in two broad categories, according to whether the energy of 
the excitation escapes from the system, or whether it is merely redistributed 
into other degrees of freedom within the system. In the first category we 
have the natural decay processes such as photon emission or particle emission.
Widths due to photon emission are completely negligible if there are any 
competing damping mechanisms, but particle emission can be important. 
The second category includes a number of diverse mechanisms for spreading 
the strength function of the oscillation, depending on the particular degrees 
of freedom which are involved. First to mention are the single-particle  
degrees of freedom. These damp the vibrational motion, a phenomenon known 
as Landau damping, if their energy spectrum is dense near the collective 
excitation energy. Another damping mechanism corresponds to collisions between 
particles within the system. Quantum-mechanically, this mechanism appears 
when configurations having a more complex character mix into the simple 
single-particle single-hole (1p-1h) 
configurations that characterize the collective 
excitations. As a rule these collisions are more effective when they take 
place at the surface of the system. This is because density is lower and the 
Pauli principle is less effective in blocking final states. Also collective
oscillations of the surface may be excited, effectively enhancing the
collisional damping.
The important role played by the surface in the purity of a nuclear vibration
is particularly simple to see in the case of non-spherical systems. 
Under such conditions, the vibrational frequency depends on the orientation
of the oscillations with respect to the axes of the system. This is
the phenomenon of inhomogeneous damping. 
Particle-decay, Landau damping and inhomogeneous damping are present already 
at the level of mean field. Collisional damping implies processes which go 
beyond mean field. The study of vibrating exotic nuclei 
in general (cf., e.g.,\ \cite{halo_gen}) and of halo nuclei 
in particular gives new possibilities to the quest for 
the mechanisms active in the damping of nuclear motion. This is because due to
the low-binding energy of the halo-nucleons, the role of damping through 
particle decay
becomes very important in these nuclei, as compared with nuclei along the 
valley of $\beta$-stability. Also because of the large radial
extension of the wavefunctions associated with the weakly bound nucleons,
new sources of inhomogeneity are active, namely those which distinguish 
between core and halo excitations. Consequently the study of the spectral
function of exotic nuclei constitutes a stringent test of theories of nuclear 
collective motion and of the associated damping processes.
\par
In the present paper we shall study the quadrupole response of $^{28}$O.
The mean field was determined within the Hartree-Fock approximation and 
the properties of the quadrupole vibrations were worked out in the Random 
Phase Approximation, renormalized through the coupling to continuum 
configurations as well as with doorway states made up of a particle-hole 
excitation and a collective surface vibration. 
It will be concluded that while the continuum plays 
an important role in determining the properties of
the linear response of the system, the corrections to 
the energy and width of the modes induced by the coupling of the vibrations to 
doorway states, not only play a role of similar importance to that played
by the coupling to the continuum, but can
change in a qualitative fashion the spectral strength 
distribution.
\par
The spherical Hartree-Fock equations have been solved using a Skyrme III (SIII)
effective interaction \cite{o_4}. 
The resulting single-particle spectrum of bound levels and the
associated mean field potential for both protons and neutrons are displayed
in Fig. 1. Also shown in Fig. 1 is the nuclear density associated with these
results. A conspicuous halo of neutron matter is apparent in keeping with the 
fact that each of the four d$_{3/2}$ neutrons of the 
system has a binding energy of 1.1 MeV. This value is to be compared
with the value of 30 MeV, that is, the binding energy of each of the least 
bound proton states, moving in the 1p$_{1/2}$ orbital. Other Skyrme
interactions (like SKM$^*$ or SGII) would give larger neutron binding 
energies and therefore, the neutron halo would be reduced or even absent. 
\par
Making use of the particle-hole basis associated with the single particle
spectrum shown above, we have calculated the corresponding self-consistent
continuum-RPA quadrupole strength function. The calculations were carried out
in coordinate space as described in \cite{Liu76}  and the corresponding
results are shown in Fig. 2.
Also shown in the figure is the unperturbed response function.
Both strength functions exhaust about 95\% of the energy weighted sum rule
of the operator $\sum_{i=1}^A r_i^2 Y_{20}({\hat r}_i)$. 
We obtain essentially the same result by solving the configuration space RPA 
matrix equation (discrete RPA)  
and by taking properly into account the coupling to the 
continuum configurations \cite{o_7}.
\par
We can roughly divide the RPA strength function in three regions :
a) The first region extends from 1.1 MeV to about 13 MeV and carries 22\%
of the EWSR. The sudden increase of the strength function at about 1.1 MeV
and the shape of the low-energy peaks is essentially controlled by the coupling 
to the continuum associated with few unperturbed single particle configurations.
In fact, we find that this part of the spectrum is dominated by transitions 
from the neutron 1d$_{3/2}$ orbital to s- and d- levels lying in the continuum.
This means that we deal with pure neutron modes, which by definition are in
equal proportion of isoscalar and isovector character. 
In keeping with these results, both the RPA and the unperturbed response
functions essentially coincide below 6 MeV; b) The second region lies in
the energy interval 13 MeV - 19 MeV and accumulates approximately 55\% of 
the EWSR. Because this part of the strength function corresponds to a single,
well defined peak, we shall identify it with the giant quadrupole resonance
(GQR) region. The fact that centroid of this resonance is at $\approx$16 MeV,
that is, lower than expected from the energy systematics of the ISGQR
($\approx 63\ A^{-1/3}$ MeV $\approx$21 MeV) and of the IVGQR 
($\approx 130\ A^{-1/3}$ MeV $\approx$43 MeV) in nuclei along the 
$\beta$-stability valley, testifies to 
the central role played by the halo nucleons
in softening this resonance. Further evidence is provided by the fact that 
the contribution of the halo neutrons to the wavefunction of states within 
this region amounts to $\approx$60\%, the contributions of neutron and proton 
core-excitations being $\approx$35\% and $\approx$5\% respectively.
Note also that in the case of nuclei along the stability valley, protons and 
neutrons contribute essentially on equal footing to the GQR; c) The third
region of the RPA-continuum strength function extends from 19 MeV up to
35 MeV and carries 
$\approx$16\% of the EWSR. Proton excitations are responsible
for $\approx$70\% of the sum rule in this region, while neutron-core 
excitations contribute with $\approx$30\%.
\par
Within the framework of the theory discussed above and making use of the 
corresponding results, we have included collisions among the nucleons
in the quadrupole strength function, that is, we have coupled the continuum-RPA
response function to vibrations of the nuclear surface. This is tantamount to 
coupling the RPA mode to ``doorway states'' (cfr. ref. [2]), containing an
uncorrelated particle-hole excitation and a collective vibration of the
nuclear surface \cite{o_9}. As can be seen from Fig. 3, a qualitative, 
let alone quantitative, change takes place in the line shape
due to doorway coupling. The most important changes are connected with the
parameters defining the giant resonance. In fact the centroid of the GQR is
lowered by $\approx$ 2 MeV, to an energy of 14 MeV, while the width of the
giant resonance is increased from its value of 4 MeV at the continuum-RPA
level, to 10 MeV. These results are model independent, and reflect the long
wavelength behaviour of the surface vibrations of halo nuclei. \par
Below 1.5 MeV two discrete peaks become evident. These results are model
dependent, being very sensitive to details of the residual interaction.
Consequently, they have to be critically interpreted, as possible although
not unique fingerprints of doorway coupling. Anyhow, the appearance of these
two low-lying peaks is due, in the present case, to the fact 
that the real part of the quadrupole vibrations self-energy arizing from 
doorway coupling lowers the energy of the transitions 1d$_{3/2} \to 
3{\rm s}_{1/2}$, 1d$_{3/2} \to 
2{\rm d}_{5/2}$, 1d$_{3/2} \to 2{\rm d}_{3/2}$ (which together exhaust about 
half of the total strength below 3 MeV), 
and brings two states based 
essentially on these configurations to an energy below particle threshold. 
The evolution of the properties of these states at different levels of the 
calculation, namely unperturbed particle-hole response, continuum-RPA, and 
continuum-RPA plus doorway coupling are displayed in Table 1. 
Particularly illuminating is the fact that the value of the imaginary part 
of the self-energy associated with the lowest states is considerably reduced 
introducing the coupling of the modes to doorway 
states, fingerprint of the lowering in energy acted on these states by the 
real part of these couplings. In the region of the GQR , doorway coupling  has
also a conspicuous effect. In particular, the centroid of the GQR is lowered
by $\approx$2 MeV to an energy of the order of 14 MeV. Furthermore, the FWHM
of the GQR changes from a value of $\approx$4 
MeV at the continuum-RPA level, to
$\approx$10 MeV including collisions.
\par
From the example of this calculation, we may conclude that in general, the
coupling to doorway states will change in a qualitative way the strength
function of vibrational states also in exotic nuclei, for which the coupling
to the continuum was expected up to now to be the main mechanism of damping.

\acknowledgments

Helpful discussions with A. Vitturi and M.A. Nagarajan are gratefully 
acknowledged.

\begin{figure}
\caption{Neutron and proton single-particle levels in $^{28}$O, obtained
within the Hartree-Fock approximation by using the effective interaction
SIII. The Hartree-Fock mean potential is also shown in the figure. Making
use of these results the density of the system has been also calculated and
is displayed by a continuous curve on top of the figure. The proton (dotted
curve) and neutron (dashed curve) contributions are also shown. }
\end{figure}

\begin{figure}
\caption{Continuum-RPA results for the quadrupole response of the nucleus 
$^{28}$O, obtained self-consistently with the interaction SIII, are
displayed with the full line. The dashed line shows the unperturbed
response. }
\end{figure}

\begin{figure}
\caption{Results of RPA plus continuum and doorway coupling for the
quadrupole response of $^{28}$O (full line). For comparison, also the
continuum-RPA results (dashed line) already shown in Fig. 2, are displayed. }
\end{figure}

\newpage

\widetext
 
\begin{table}
\caption{Evolution of the energy of the two sharp states in the very low 
energy region of the spectrum. Results associated with the unperturbed
discrete  
response, continuum-RPA and
continuum-RPA plus doorway coupling are shown in the upper, middle and  
lower section of the table. } 
\vspace{0.4cm}
\begin{tabular}{ccc} 
\multicolumn{3}{c}{Unperturbed response} \\ \tableline
Transition Energy [MeV]&Energy [MeV]  &Percentage of  Strength \\ \tableline
1d$_{3/2}\to 3{\rm s}_{1/2}$ & 1.60 & 3.3\\
1d$_{3/2}\to 2{\rm d}_{5/2}$ & 2.17 & 1.3\\
1d$_{3/2}\to 2{\rm d}_{3/2}$ & 2.19 & 2.2\\ \tableline
\multicolumn{3}{c}{Continuum-RPA} \\ \tableline 
1d$_{3/2}\to 3{\rm s}_{1/2}$ & 1.31-i0.18 & 1.8\\
1d$_{3/2}\to 2{\rm d}_{5/2}$ & 1.82-i0.34 & 1.0\\
1d$_{3/2}\to 2{\rm d}_{3/2}$ & 1.82-i0.34 & 0.7\\ \tableline
\multicolumn{3}{c}{Continuum-RPA plus doorway coupling} \\ \tableline 
1d$_{3/2}\to 3{\rm s}_{1/2}$ & 0.78-i0.01 & 2.3\\
1d$_{3/2}\to 2{\rm d}_{5/2}$ & 1.15-i0.02 & 0.6\\
1d$_{3/2}\to 2{\rm d}_{3/2}$ & 1.15-i0.02 & 1.1\\ 
\tableline
\end{tabular}
\end{table}


\begin{references}
\bibitem{o_4} The calculations have been carried out as in ref. [6] to
              which we refer the reader for details except that we now
              calculate the mean field unoccupied levels at positive energies
              by putting the system inside a box 
              of radius 25 fm instead of diagonalizing
              the mean field on a harmonic oscillator basis.
\bibitem{o_7}   If we do not include the coupling to the continuum we obtain
                the results of ref. [9].
\bibitem{o_9}   The collective vibrations included are the 
                isoscalar 2$^+$, 3$^-$ and the isovector 1$^-$ states with 
                more than 1\% of the total strength and energy below 10 MeV, 
                and with more than 2\% of the total strength and energy 
                between 10 and 20 MeV.
\end{references}
\end{document}